\documentstyle[aps,twocolumn,epsf,tighten]{revtex}
\newcommand{\tr}{\mbox{tr}\,}
\newcommand{\Tr}{\mbox{Tr}\,}
\newcommand{\re}{\mbox{Re}\,}
\begin{document}
\draft
\title
{\hfill\begin{minipage}{0pt}\scriptsize \begin{tabbing}
\hspace*{\fill} GUTPA/00/06/02\end{tabbing} 
\end{minipage}\\[8pt]  
Casimir scaling of SU(3) static potentials}
\author{Gunnar Singh Bali\thanks{Electronic
address: g.bali@physics.gla.ac.uk},}
\address{Department of Physics and Astronomy, University of Glasgow,
Glasgow G12 8QQ, Scotland}
\date{\today}
\maketitle

\begin{abstract}
Potentials between static colour sources in eight different representations
are computed in four dimensional
$SU(3)$ gauge theory. The simulations have been
performed with the
Wilson action on anisotropic lattices where the renormalised
anisotropies have been determined non-perturbatively. After an
extrapolation to the continuum limit we are able to
exclude any violations
of the Casimir scaling hypothesis that exceed 5~\%
for source separations of up to 1~fm.

\end{abstract}

\narrowtext
\section{Introduction}
Non-perturbative QCD effects in general and the nature of
the confinement mechanism in particular are theoretically challenging.
At the same time these aspects are important for high energy
and low energy particle and nuclear phenomenology.
Several models of non-perturbative QCD have been
proposed whose predictions happen to differ from each
other substantially in some cases. Prominent examples are bag
models~\cite{Chodos:1974je,DeGrand:1975cf,Hasenfratz:1978dt,DeTar:1983rw,Juge:1998nd},
strong coupling and flux tube
models~\cite{Kogut:1976zr,Buchmuller:1982fr,Isgur:1983wj,Isgur:1985bm},
bosonic string models~\cite{Goddard:1973qh,Luscher:1981iy},
the
stochastic vacuum model~\cite{Dosch:1987sk,Dosch:1988ha,Simonov:1988rn},
dual QCD~\cite{Baker:1983bt,Baker:1995nq,Baker:1996mk},
the Abelian Higgs model~\cite{Maedan:1988yi,Chernodub:1999xf},
and instanton based
models~\cite{Shuryak:1982ff,Diakonov:1986eg,Schafer:1998wv}.
Lattice simulations of interactions between static colour
sources offer an ideal environment for discriminating between
different models of low energy QCD and to learn more about the
confinement mechanism. They are easily accessible analytically
and at the same time very accurate Monte Carlo predictions can be
obtained~\cite{Bali:1992ab,Bali:1993ru}.

Despite the availability of
a wealth of information on fundamental potentials,
only few lattice investigations of forces 
between sources in higher
representations of $SU(N)$ gauge groups exist.
Most of these studies have been performed in $SU(2)$ gauge theory in
three~\cite{Ambjorn:1984dp,Poulis:1997nn,Stephenson:1999kh,Philipsen:1999wf}
and
four~\cite{Bernard:1982pg,Bernard:1983my,Ambjorn:1984mb,Michael:1985ne,Griffiths:1985ip,Trottier:1995fx,deForcrand:1999kr,Kallio:2000jc}
space-time dimensions.
Zero temperature results for four dimensional $SU(3)$ can be found in
Refs.~\cite{Campbell:1986kp,Michael:1992nc,Michael:1998sm,Deldar:1998ne,Bali:2000hx,Deldar:1999vi}
while determinations of Polyakov line correlators in
non-fundamental representation
have been performed at finite temperature by
Bernard~\cite{Bernard:1982pg,Bernard:1983my} for $SU(2)$
and in Refs.~\cite{Ohta:1986pc,Markum:1988na,Muller:1991xj,Buerger:1993bq}
for $SU(3)$ gauge
theory.

In our study we shall see that the so-called Casimir scaling
hypothesis~\cite{Ambjorn:1984dp}
is rather accurately represented by the lattice data
while models predicting a different behaviour are
definitely ruled out. Casimir scaling means that potentials
between sources in different representations are proportional
to each other with their ratios given by the respective
ratios of the  eigenvalues of the corresponding quadratic
Casimir operators, which is exact in the case of two dimensional
Yang-Mills theories.
Our result is of particular interest with respect
to recent discussions of the confinement
scenario~\cite{Faber:1998rp,Cornwall:1998ds,Deldar:1999yy,Simonov:2000sw,Shevchenko:2000du}.

At distances $r>r_c\approx 1.2$~fm~\cite{Michael:1998sm} non-fundamental
sources will be screened and ``string breaking'' effects will be encountered
that are incompatible with Casimir scaling. In the present
study we restrict ourselves
to distances smaller than the string breaking scale $r_c$.

This article is organised as follows: in Sec.~\ref{sec:extract}
the lattice methods that we apply and our notations are introduced.
A determination of the renormalised anisotropies and
lattice spacings is presented in
Sec.~\ref{sec:aniso}. The potentials are then
determined in Sec.~\ref{sec:pots} before we conclude with
a brief discussion.

\section{Notations and Methods}
\label{sec:extract}
We denote the energy of colour sources, separated by a distance
$r$, in a representation
$D={\mathbf 3},{\mathbf 6},{\mathbf 8},{\mathbf 1}{\mathbf 0},\ldots$
of the $SU(3)$ gauge group by $V_D(r,\mu)$, where $\mu$ denotes some
cut-off scale on the gluon momenta, for instance an inverse
lattice spacing, $\mu=\pi/a$.
We shall also use the subscript ``$F$'' to label the fundamental
(${\mathbf 3}$) representation (or we may just omit the subscript
in this case).

\begin{table}[hbt]
\caption{Group theoretical factors for $SU(3)$. $D$ is the dimension of the
  representation, $(p,q)$ are the weight factors, $z=\exp(2\pi i/3)$,
  and $d_D=C_D/C_F$ denote ratios of quadratic Casimir charges.}
\label{tab:reps}
\begin{center}
\begin{tabular}{c|r|c|c|l}
$D$&$(p,q)$&$z^{p-q}$&$p+q$&$d_D$\\\hline
3&$(1,0)$&$z$&1&1\\
8&$(1,1)$&1&2&2.25\\
6&$(2,0)$&$z^*$&2&2.5\\
$15a$&$(2,1)$&$z$&3&4\\
10&$(3,0)$&1&3&4.5\\
27&$(2,2)$&1&4&6\\
24&$(3,1)$&$z^*$&4&6.25\\
$15s$&$(4,0)$&$z$&4&7
\end{tabular}
\end{center}
\end{table}

The static potential,
\begin{equation}
\label{eq:po1}
V_D(r,\mu)=V_D(r)+V_{D,\mbox{\scriptsize self}}(\mu),
\end{equation}
can be factorised into
an interaction part, $V_D(r)$, and a self energy contribution,
$V_{D,\mbox{\scriptsize self}}(\mu)$ that will diverge like $\mu/\ln\mu$
as $\mu\rightarrow\infty$ while $V_D(r)$ will assume universal values.

A (dimensionless) lattice potential $\hat{V}_D({\mathbf R},a)$
will resemble the corresponding
continuum potential up to lattice artefacts,
\begin{eqnarray}
\label{eq:conte}
V_D(Ra)&=&a^{-1}\left[\hat{V}_D({\mathbf R},a)-
\hat{V}_{D,\mbox{\scriptsize self}}(a)\right]\\\nonumber
&\times&\left[1
+f_D(Ra,\hat{\mathbf R})a^{\nu}\right],
\end{eqnarray}
where $\nu$ is a positive integer number that will in general depend
on the lattice action employed. We are concerned
with Wilson-type gluonic actions~\cite{Wilson:1974sk}.
In this case, $\nu=2$. Note that the coefficient function $f$
only depends on the combination
$r=Ra$ and on the direction of ${\mathbf R}$
but not on $R$ itself. This guarantees that lattice artefacts
are reduced as $r\gg a$ is increased.

The static potentials are obtained from fits to
smeared~\cite{Albanese:1987ds,Bali:1992ab,Bali:1995de}
Wilson loops for $T\geq T_{\min}$ where $T_{\min}$
depends on ${\mathbf R}$,  the statistical errors of the Wilson loops
and the smearing algorithm employed. We define a Wilson
loop in representation $D$,
\begin{equation}
\label{eq:wil}
W_D({\mathbf R},T)=
\Tr \left(\prod_{(n,\mu)\in\delta C({\mathbf R},T)}U_{D,n,\mu}\right),
\end{equation}
where $\delta C({\mathbf R},T)$ denotes the oriented boundary of
a (generalised) rectangle with spatial extent ${\mathbf R}$ and a
temporal separation of $T$ lattice spacings.
$(n,\mu)$ denotes an oriented link connecting
the site $n$ with $n+\hat{\mu}$, $n$ is an integer four-vector
that labels a lattice site and $\hat{\mu}$ is a unit vector
pointing into a direction, $\mu\in\{1,2,3,4\}$.
``$\Tr $'' is the
normalised trace, $\mbox{Tr}_D{\mathbf 1}_D=\frac{1}{N_D}\tr 
{\mathbf 1}_D=1$, $N_D$ is the dimension of the representation $D$ and,
\begin{equation}
U_{D,n,\mu}={\mathcal P}\left\{\exp\left[
i\int_{an}^{a(n+\hat{\mu})}\!dx_{\mu}\,A^a_{\mu}(x)T_a^D
\right]\right\},
\end{equation}
denotes a link variable in representation $D$, where
$T_a^D$ is a generator of the gauge group. Our conventions are
$[T_a^D,T_b^D]=if_{abc}T_c^D$, where $f_{abc}$ are totally antisymmetric
real structure constants. The normalisation is such that
$\tr T_a^FT_b^F=\delta_{ab}/2$.
Now:
\begin{equation}
\label{eq:po2}
\langle W_D({\mathbf R},T)\rangle=c_D({\mathbf R},a)
\exp\left[-\hat{V}_D({\mathbf R},a)T\right]
\quad(T\rightarrow\infty).
\end{equation}

The use of smeared Wilson loops turns out to be
more suitable for numerical simulations than implementing the definition
of Eq.~(\ref{eq:wil}); the spatial pieces of the Wilson loop
are replaced by linear combinations of various paths that models the
ground state wave function. As a result
the overlap of the creation operator with this ground state, $c_D$,
is enhanced
and the $T\rightarrow\infty$ limit can effectively be realised
at moderate $T$ values.

In numerical simulations one observes that the statistical
error $\Delta W(T)$ of the expectation value of a smeared
Wilson loop $\langle W(T)\rangle$ only weakly varies with
$T$~\cite{Bali:1995de}.
From Eqs.~(\ref{eq:po1}) and (\ref{eq:po2}) we therefore obtain
the relation,
\begin{equation}
\label{eq:errorr}
\frac{\Delta W(T)}{\langle W(T)\rangle}\propto\exp[\hat{V}(a)T]\propto
\exp\left[\hat{V}_{\mbox{\scriptsize self}}(a)T\right]
\quad(T\rightarrow\infty),
\end{equation}
for the relative errors (which are directly
proportional to the statistical uncertainty of the potential values).
In tree level perturbation theory one finds,
\begin{equation}
\label{eq:selfen}
\hat{V}_{D,\mbox{\scriptsize self}}(a)=c\,C_D\,g^2(a), \quad
c=0.252731\ldots;
\end{equation}
the self energy is proportional to the eigenvalue of the quadratic Casimir
operator $C_D=\mbox{Tr}_DT_a^DT_a^D$ of the representation.
This means that statistical errors will increase significantly
as we investigate higher representations of the sources with
bigger Casimir charges.

This self energy related problem motivates us to
introduce an anisotropy parameter $\xi=a_{\sigma}/a_{\tau}\approx 4$
between spatial lattice resolution $a_{\sigma}$ and temporal
lattice constant $a_{\tau}$. This results
in a reduction of the self energy,
$\hat{V}_{D,\mbox{\scriptsize self}}= c\,C_D\,\xi^{-1}g^2$,
and therefore of the relative errors of smeared
Wilson loops. However, at the end of the day
we wish to measure distance and potential in the same 
units $a_{\sigma}$. This means that we do not gain anything from the
factor $\xi$ within the above expression.
Nevertheless, a $\xi>1$ still results in a
reduced effective $g^2$ at fixed $a_{\sigma}$. Of course, equally well
we could just have increased the size of our statistical
ensemble by a factor four and worked at $\xi=1$.

Our main motivation for introducing an anisotropy is the possibility
of reducing lattice artefacts.
These are most prominent at small and at large distances:
as long as $r$ is not much larger than $a_{\sigma}$ the cubic lattice
structure is clearly visible. In lowest order perturbation theory
these violations of rotational symmetry
only depend on ${\mathbf R}$ and $a_{\sigma}$
while the order $g^4$ coefficients exhibit a
weak dependence on $\xi$ too~\cite{Altevogt:1995cj}.
While we cannot hope to significantly reduce these small
distance effects without decreasing $a_{\sigma}$, it is clear that
on a lattice with temporal resolution $a_{\tau}$ one cannot
reliably resolve masses
$m\gg a_{\tau}^{-1}$.
However, at $r\gg a_{\sigma}$ and in particular for representations
$D$ with large Casimir charges
situations, $V_D(r)a_{\sigma}\gg 1$, are easily
encountered, unless $\xi\gg 1$.

Introducing an anisotropy also means that within any physical $t$
window we have more data points at our disposal. While this might
help to gain more confidence in identifying effective mass plateaus
we find that the additional data points are highly correlated
and add little extra information, at least when one is only
interested in the mass of the ground state within a given channel.

Our action reads,
\begin{equation}
\label{eq:act}
S=-\beta\sum_n\left[\frac{1}{\xi_0}\sum_{i>j}\Tr  U_{n,ij}
+\xi_0\sum_i\Tr U_{n,i4}\right],
\end{equation}
where $\beta=2N/[g^2(a_{\sigma},\xi)]$ is defined through
the lattice coupling $g^2$ and $i,j\in\{1,2,3\}$. $U_{n,\mu\nu}=
U_{n,\mu}U_{n+\hat{\mu},\nu}U^{\dagger}_{n+\hat{\nu},\mu}
U^{\dagger}_{n,\nu}$ denotes the product of four link variables
around an elementary square, the plaquette.
With the above anisotropic Wilson action, the leading order lattice artefacts
are proportional to $a_{\sigma}^2$
and $a_{\tau}a_{\sigma}=a_{\sigma}^2\xi^{-1}$. This means that along a
trajectory of constant $\xi$ the continuum limit will be approached
quadratically in $a_{\sigma}$. The relationship between
the bare anisotropy $\xi_0$ appearing in the action Eq.~(\ref{eq:act})
and the renormalised anisotropy $\xi$
is known in one loop perturbation theory~\cite{Karsch:1982ve}:
$\xi=\xi_0[1+c_1(\xi_0)g^2+\cdots]$.
In Sec.~\ref{sec:aniso} we will discuss
our non-perturbative evaluation of $\xi$. Of course the function
$\xi(g^2,\xi_0)$
is not unique and different non-perturbative definitions
might differ from each other by terms of order $a_{\sigma}$.

Perturbation theory yields the relation between potentials
in different representations $D$,
\begin{equation}
\label{eq:pothigh}
V_D(r,\mu)=d_DV_F(r,\mu),
\end{equation}
where $d_D=C_D/C_F$.
Table~\ref{tab:reps} contains all representations $D$,
the corresponding weights $(p,q)$ and
the ratios of Casimir factors, $d_D$, for $p+q\leq 4$.
In $SU(3)$ we have $C_F=4/3$,
and $z=\exp(2\pi i/3)$ denotes a third root of 1.
Eq.~(\ref{eq:pothigh}) is known to hold to (at least) one loop
(order $g^4$) perturbation theory
at finite lattice spacing~\cite{Weisz:1984bn}
and two loops (order $g^6$)
in the continuum limit~\cite{Schroder:1999vy}
of four dimensional Yang-Mills theories.
The main purpose of this article is to investigate non-perturbatively
to what extent the ``Casimir scaling''
relation Eq.~(\ref{eq:pothigh}) is violated.

In what follows $U$ denotes a group element in the fundamental
representation of $SU(3)$, for instance the product of link
variables around a closed contour. 
The traces of $U_D$ in various representations,
$V_D=\tr U_D$, can easily be expressed in terms of traces
of powers of $U$,
\begin{eqnarray}
\label{eq:reps}
V_3&=&\tr U,\\
V_8&=&\left(|V_3|^2-1\right),
\\
V_6&=&\frac{1}{2}\left[(\tr U)^2+\tr U^2\right],
\\
V_{15a}&=&\tr U^*\,V_6-\tr U,\\
V_{10}&=&\frac{1}{6}\left[(\tr U)^3+3\,\tr U\,\tr U^2+2\,\tr U^3\right],\\
V_{24}&=&\tr U^*\,V_{10}-V_6,\\
V_{27}&=&|V_6|^2-|V_3|^2,\\
V_{15s}&=&\frac{1}{24}\left[
(\tr U)^4+6(\tr U)^2\tr U^2
+3(\tr U^2)^2\right.\nonumber\\
&+&\left. 8\,\tr U\,\tr U^3+6\,\tr U^4\right].
\label{eq:reps15s}
\end{eqnarray}
Note that $\mbox{Tr}_DU_D=\frac{1}{N_D}\tr U_D=V_D/N_D$.
Hence, the normalisation of $V_D$ differs by a factor $N_D$
from that of the Wilson loop $W_D$ of Eq.~(\ref{eq:wil}).
Under the replacement, $U\rightarrow z\,U$, $W_D$ transforms like,
$W_D\rightarrow z^{p-q}W_D$. Representations with $z^{p-q}=1$
have zero triality.

\section{Determining anisotropy and lattice spacing}
\label{sec:aniso}
We simulate $SU(3)$ gauge theory at the parameter values
$(\beta,\xi_0)=(5.8,3.10), (6.0,3.20)$ and $(6.2,3.25)$.
{}From exploratory simulations with limited statistics
and the published data
of Refs.~\cite{Klassen:1998ua,Ejiri:1998xd,Engels:2000tk}
we expect to find renormalised anisotropies
$\xi\approx 4$ at these combinations.
At the above parameter values volumes of $L_{\sigma}^3\times L_{\tau}=
8^3\times 48$,
$12^3\times 72$ and $16^3\times 96$ lattice sites have been
realised, respectively. These volumes were chosen to keep the
lattice extent about constant in physical units. In addition
a volume of $12^3\times 48$ lattice sites has been simulated
at $\beta=5.8$ to investigate possible finite
size effects.

The gauge configurations have been obtained
by randomly mixing Cabibbo-Marinari style~\cite{Cabibbo:1982zn}
Fabricius-Haan heatbath~\cite{Fabricius:1984wp}
and Creutz overrelaxation~\cite{Creutz:1987xi}
sweeps, where we cycled over the
three diagonal $SU(2)$ subgroups. The probability of a heatbath sweep
was set to be $1/5$. During each sweep the sites were visited
subsequently for each of the four space-time directions of the links
in sequential order. Measurements were taken
after 2000 initial heatbath sweeps and the
gauge configurations are separated
from each other
by 200 sweeps. In doing so, we did not find any signs of autocorrelation
or thermalization effects for any of the investigated observables at
any of the simulated 
parameter sets. In the case of $\beta=6.0$ one set of configurations
was generated on a Sparc station after a cold start while another set
of configurations was generated on a Cray J90, starting from
a hot, random configuration. No statistically significant deviations
between these two data sets were found either. We display our simulation
parameters in Table~\ref{tab:r02}. $n_{\mbox{\scriptsize conf}}$ denotes
the number of statistically independent configurations analysed in each
case while $r_0\approx 0.5$~fm is the Sommer scale
parameter~\cite{Sommer:1994ce},
implicitly
defined through the static potential,
\begin{equation}
\label{eq:sommer}
\left.\frac{dV(r)}{dr}\right|_{r=r_0}=1.65.
\end{equation}

\begin{table}[hbt]
\caption{Simulation parameters,
lattice spacings and linear lattice extents.}
\label{tab:r02}
\begin{center}
\begin{tabular}{c|c||c|c|c|c}
$\beta$&$L_{\sigma}$&$n_{\mbox{\scriptsize conf}}$&
$r_0/a_{\sigma}$&
$r_0/a_{\tau}$&$L_{\sigma}a_{\sigma}/r_0$\\\hline
5.8&8 &633&3.074(29)&12.45(11)&2.60(3)\\
5.8&12&139&3.052(34)&12.52(13)&3.93(4)\\
6.0&12&159&4.437(71)&18.12(24)&2.70(5)\\
6.2&16&133&6.106(66)&24.16(23)&2.62(3)
\end{tabular}
\end{center}
\end{table}

\begin{figure}
\centerline{\epsfxsize=8cm\epsfbox{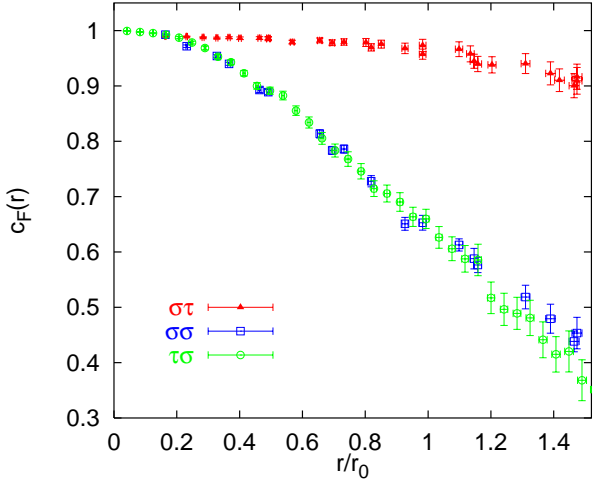}}
\caption{Ground state overlaps for the case of the fundamental potential
determined from Wilson loops in three different orientations at
$\beta=6.2$, $\xi_0=3.25$.}
\label{fig:over}
\end{figure}

We label quantities associated with the fine grained direction
by an index $\tau$ while $\sigma$ refers to coarse grained directions.
On an anisotropic lattice various ways of
associating the sides of smeared Wilson loops
with these directions exist:
$W_{\tau\sigma}(r/a_{\tau},t/a_{\sigma})$,
$W_{\sigma\sigma}({\mathbf r}/a_{\sigma},t/a_{\sigma})$ and
$W_{\sigma\tau}({\mathbf r}/a_{\sigma},t/a_{\tau})$.
In the case of $W_{\tau\sigma}$ as well as for $W_{\sigma\sigma}$ the time
coordinate points into a $\sigma$ direction. The spatial
coordinate is identified with the $\tau$ direction in the first case
and a $\sigma$ direction in the latter case. While the spatial
connections within $W_{\tau\sigma}$ are parallel to the $\tau$ axis,
in the case of $W_{\sigma\sigma}$ we realise
planar off-axis configurations ${\mathbf R}\parallel (1,1,0)$
and ${\mathbf R}\parallel (2,1,0)$,
in addition to on-axis separations\footnote{
We ignore the possibility of mixing
$\tau$ and $\sigma$ coordinates within the ``spatial'' separation.}.
Finally, within $W_{\sigma\tau}$ the time coordinate is taken along
the fine grained dimension and the spatial coordinate coarse grained.
We determine $W_{\sigma\tau}$ for the standard separations~\cite{Bali:1992ab},
${\mathbf R}\parallel (1,0,0),$ $(1,1,0),$ $(2,1,0),$ $(1,1,1),$ $(2,1,1),$
$(2,2,1)$.

In the spatially isotropic situation ($W_{\sigma\tau}$)
we iteratively construct fat links in the standard way
by replacing a given link by the sum of itself and the neighbouring
four spatial staples with some weight parameter, $\alpha\geq 1$,
\begin{eqnarray}
\label{eq:smear}
U_{n,i}&\rightarrow&P_{SU(3)}\left(\alpha\,U_{n,i}+
\sum_{j\neq i}F_{n,j}\right),\\\nonumber
F_{n,j}&=&U_{n,j}U_{n+\hat{\jmath},i}U^{\dagger}_{n+\hat{\imath},j}+
U_{n-\hat{\imath},i}^{\dagger}
U_{n-\hat{\imath},i}U_{n-\hat{\imath}+\hat{\jmath},j}.
\end{eqnarray}
$P_{SU(3)}$ denotes a projection operator, back onto the $SU(3)$ manifold.
We employ the definition~\cite{Bali:1992ab}, $U=P_{SU(3)}(A)\in SU(3)$,
$\re\Tr  UA^{\dagger}=\max$ and iterate Eq.~(\ref{eq:smear})
26 times with $\alpha=2.3$.

We use a somewhat different novel smearing algorithm
in the case of $W_{\tau\sigma}$ and $W_{\sigma\sigma}$
where the
spatial volume is anisotropic with one fine and two coarse
directions: when one only considers links parallel to the one being replaced,
Eq.~(\ref{eq:smear}) resembles a two dimensional diffusion process:
$U\rightarrow P_{SU(3)}[(\alpha+4+\nabla^2_2)U]$.
We are interested to maintain an isotropic propagation of the link fields
when an anisotropy parameter $\xi>1$ is introduced. Following
the above diffusion model this is achieved
by replacing Eq.~(\ref{eq:smear}) with,
\begin{eqnarray}
\label{eq:smear2}
U_{n,i}&\rightarrow& P_{SU(3)}\left(\alpha\, U_{n,i}+ F_{n,j}+
\xi^2F_{n,\tau}\right),\\\label{eq:smear3}
U_{n,\tau}&\rightarrow& P_{SU(3)}\left[(\alpha+2\xi^2-2) U_{n,\tau}+
\sum_iF_{n,i}\right],
\end{eqnarray}
where $i,j\in\{1,2\}, j\neq i$. We perform 22 iterations of
Eqs.~(\ref{eq:smear2}) -- (\ref{eq:smear3})
with $\alpha=3.7$. Indeed, in doing so we find very similar overlaps
with the physical ground state, $c_{\sigma\sigma}({\mathbf R})\approx
c_{\tau\sigma}({\mathbf R})$, where $c=c_F\in[0,1]$
is defined in Eq.~(\ref{eq:po2})
and $W_{\tau\sigma}({\mathbf R},T=0)=W_{\sigma\sigma}({\mathbf R},T=0)=1$.
This is illustrated in the comparison of data obtained
at $\beta=6.2$, $\xi_0=3.25$ of Fig.~\ref{fig:over}. We have
not been able, however, to sustain the high overlaps achieved
for $W_{\sigma\tau}$ (triangles)
for $W_{\tau\sigma}$ or $W_{\sigma\sigma}$.
The situation at the other $(\beta,\xi_0)$ combinations is similar.

From the asymptotic behaviour of the different Wilson loops at large
temporal separations $t$, three lattice potentials can be determined:
\begin{eqnarray}
\label{eq:effm1}
a_{\sigma}^{-1}\hat{V}_{\tau\sigma}(r/a_{\tau})&=&
-\lim_{t\rightarrow\infty}
\frac{d}{dt}\ln
W_{\tau\sigma}(r/a_{\tau},t/a_{\sigma}),\\\label{eq:effm2}
a_{\sigma}^{-1}\hat{V}_{\sigma\sigma}({\mathbf r}/a_{\sigma})&=&
-\lim_{t\rightarrow\infty}
\frac{d}{dt}\ln
W_{\sigma\sigma}({\mathbf r}/a_{\sigma},t/a_{\sigma}),\\
a_{\tau}^{-1}\hat{V}_{\sigma\tau}({\mathbf r}/a_{\sigma})&=&
-\lim_{t\rightarrow\infty}
\frac{d}{dt}\ln
W_{\sigma\tau}({\mathbf r}/a_{\sigma},t/a_{\tau}).
\end{eqnarray}
Note that while $\hat{V}_{\tau\sigma}$
and $\hat{V}_{\sigma\sigma}$ are given in units of $a_{\sigma}$,
$\hat{V}_{\sigma\tau}$ is measured in units of $a_{\tau}$.
These potentials are related to each other:
\begin{eqnarray}
\label{eq:match}
\hat{V}_{\sigma\sigma}(R,a_{\sigma})&=&
\hat{V}_{\tau\sigma}(\xi R,a_{\sigma})[1+{\mathcal
O}(a_{\sigma})^2]\\\label{eq:match2}
&=&\left[\xi\hat{V}_{\sigma\tau}(R,a_{\sigma})+\Delta
\hat{V}_{\mbox{\scriptsize self}}(a_{\sigma})\right]
[1+{\mathcal O}(a_{\sigma})^2],
\end{eqnarray}
where
$\Delta\hat{V}_{\mbox{\scriptsize self}}=  
\hat{V}_{\sigma\sigma,\mbox{\scriptsize self}}-\xi
\hat{V}_{\sigma\tau,\mbox{\scriptsize self}}$.
While $\hat{V}_{\sigma\sigma}$ and $\hat{V}_{\tau\sigma}$ are equal
at a given physical distance (up to lattice artefacts),
in the case of $\hat{V}_{\sigma\tau}$ a shift by an additive constant
is expected since the
self energies differ:
\begin{equation}
\label{eq:vdiff}
\Delta\hat{V}_{\mbox{\scriptsize self}}(a_{\sigma})=
0.08214\ldots g^2+\cdots.
\end{equation}
The numerical value has been obtained
in lowest order perturbation theory for $\xi=\xi_0=4$.

Following Ref.~\cite{Klassen:1998ua} we use
Eq.~(\ref{eq:match}) to determine
the renormalised anisotropies\footnote{Note that unlike
Ref.~\cite{Klassen:1998ua}
our analysis is based on asymptotic $T\rightarrow\infty$ results rather
than on pre-asymptotic finite $T$ approximants to the potential.}.
Eq.~(\ref{eq:match2}) can then be
used as an independent consistency check (modulo
lattice artefacts).
In order to guarantee a consistent
definition of $\xi$
we can either consider the limit~\cite{Engels:2000tk} $R\rightarrow\infty$ or
demand the matching to be performed at the same distance
in terms of a measured correlation length. We follow
the latter strategy and impose,
\begin{equation}
\label{eq:impose}
\hat{V}_{\sigma\sigma}(R_m^L,a_{\sigma})=
\hat{V}_{\tau\sigma}(\xi R_m^L,a_{\sigma}),
\end{equation}
at $R_m^L\approx R_m= (2/3) r_0/a_{\sigma}$
where $r_0\approx 0.5$~fm is the Sommer scale of Eq.~(\ref{eq:sommer}).
We restrict ourselves to
on-axis separations and take $R_m^L=2,3,4$ for $\beta=5.8,6.0$ and
$6.2$, respectively. This choice is justified by our subsequent
analysis where we find
$R_m=2.05(2)$ and $R_m=2.03(2)$ on the $8^3$ and $12^3$
$\beta=5.8$ lattices and
$R_m=2.96(4)$ and $R_m=4.07(4)$ at $\beta=6.0$ and $\beta=6.2$,
respectively. The renormalised anisotropy $\xi$ is then obtained from
Eq.~(\ref{eq:impose}) by
interpolating the potential $V_{\tau\sigma}$
according to three parameter fits,
\begin{equation}
\label{eq:fitp}
\hat{V}_{\tau\sigma}(R)=\hat{V}_{0,{\tau\sigma}}+K_{\tau\sigma}\,R
-\frac{e_{\tau\sigma}}{R}.
\end{equation}
The errors are obtained via the bootstrap procedure.
The resulting anisotropies and fit ranges employed,
$R\in [R_{\min},L_{\tau}/2]$, as well as the fit parameters in
units of $a_{\sigma}$ are displayed in Table~\ref{tab:aniso}.
$T_{\min}=t_{\min}/a_{\sigma}$ denotes the ``temporal'' separation from
which onwards effective mass data [Eqs.~(\ref{eq:effm1})
and (\ref{eq:effm2})] saturated into plateaus.

The renormalised anisotropies $\xi$ are also included in the
last column of\footnote{Note that in this table we have averaged
the results obtained on the $8^3$ and $12^3$ lattices at
$\beta=5.8$ that agree with each other within errors.}
Table~\ref{tab:aniso2}. In the third column of this table
the one loop results~\cite{Karsch:1982ve,GarciaPerez:1997ft}
are displayed while in
the second last column mean field
estimates~\cite{Parisi:1980pe,Lepage:1993xa}
are shown,
\begin{equation}
\label{eq:tadpole}
\xi_{ir}=\xi_0\sqrt{\frac{\langle U_{\sigma\tau}\rangle}
{\langle U_{\sigma\sigma}\rangle}},\quad
\beta_{ir}=\beta\sqrt{\langle U_{\sigma\tau}\rangle
\langle U_{\sigma\sigma}\rangle}.
\end{equation}
The temporal and spatial average plaquettes ($\langle U_{\sigma\tau}\rangle$
and $\langle U_{\sigma\sigma}\rangle$) are also included in the table.
While the renormalised anisotropies are underestimated
by the one loop results by about 10~\% they are overestimated
by the mean field values by almost the same amount.
Finally, in Table~\ref{tab:aniso3} we compile the bare anisotropies
$\xi_{0,4}=4\,\xi_0/\xi$
at which we should have simulated in order to achieve $\xi=4$.
In doing so we assume
that our statistical uncertainties
on $\xi$ of order 1~\% will dominate over
variations of the ratios $\xi_0/\xi$
under a change of $\xi_0$ by less than 2~\%.

After determining the anisotropies, the
potential
$V_{\sigma\tau}$ is fitted to the parametrisation
Eq.~(\ref{eq:fitp}) for $r\geq r_m^L\approx r_m=2r_0/3$.
The results are compiled in Table~\ref{tab:r0}.
The fit parameters $e$ and $K$ agree within errors with those determined
from the data on $V_{\tau\sigma}$ of
Table~\ref{tab:aniso} while the $V_0$ values tend to be somewhat smaller,
in agreement with the expectation of Eq.~(\ref{eq:vdiff}).
Note that the parametrisation is thought to be effective only and that
the fit ranges employed for the two potentials differ from each other.
{}From the fit parameters,
values $r_0/a_\sigma=\sqrt{(1.65\,\xi^{-1}-e)/K}$
can be extracted. These are displayed in Table~\ref{tab:r02}, along with
the linear spatial lattice extent. Compared to the isotropic case,
$\xi_0=1$, where~\cite{Schilling:1993bk,Bali:2000gf}
$r_0/a=3.64(5), 5.33(3)$ and $7.29(4)$ at $\beta=5.8,6.0$ and $6.2$,
respectively, $a_{\sigma}$ is somewhat
increased while the temporal lattice spacing $a_{\tau}$ is
reduced. The ratios
$r_0a_{\sigma}^{-3/4}a_{\tau}^{-1/4}=4.36(4),$ $6.31(10)$
and $8.61(9)$ exhibit that at $\xi=4$
the geometrically averaged lattice spacings
are about 15~\% smaller than their isotropic counterparts,
obtained at the same $\beta$ values.

\begin{table}[hbt]
\caption{Fits to $V_{\tau\sigma}$ and determination of the renormalised
anisotropy $\xi$. The fit parameters are displayed in units of $a_{\sigma}$.}
\label{tab:aniso}
\begin{center}
\begin{tabular}{c|c||c|c|c|c|c|c}
$\beta$&$L_{\sigma}$&$\frac{t_{\min}}{a_{\sigma}}$&$\frac{r_{\min}}{a_{\tau}}$&
$\xi$&$\hat{V}_{0,\tau\sigma}$&$e_{\tau\sigma}\xi^{-1}$&
$K_{\tau\sigma}\xi$\\\hline
5.8&8 &3&6&4.052(32)&0.826(43)&0.352(43)&0.143(11)\\
5.8&12&3&6&4.102(33)&0.834(45)&0.366(44)&0.144(12)\\
6.0&12&4&5&4.084(74)&0.813(20)&0.332(18)&0.0697(49)\\
6.2&16&4&8&3.957(51)&0.792(18)&0.356(28)&0.0384(28)
\end{tabular}
\end{center}
\end{table}

\begin{table}[hbt]
\caption{Bare, one loop, and mean field estimated
(``tadpole improved'')
anisotropies versus the non-perturbatively determined renormalised
anisotropy.}
\label{tab:aniso2}
\begin{center}
\begin{tabular}{c|c||c|c|c|c|c}
$\beta$&$\xi_0$&$\xi_{g^2}$&$\langle U_{\sigma\sigma}\rangle$&
$\langle U_{\sigma\tau}\rangle$&$\xi_{ir}$&$\xi$\\\hline
5.8&3.10&3.470&0.41557(3)&0.807175(5)&4.320&4.076(25)\\
6.0&3.20&3.573&0.44540(2)&0.820463(4)&4.343&4.084(74)\\
6.2&3.25&3.619&0.46924(1)&0.829761(2)&4.322&3.957(51)
\end{tabular}
\end{center}
\end{table}

\begin{table}[hbt]
\caption{Estimates of the bare anisotropies that correspond to $\xi=4$.}
\label{tab:aniso3}
\begin{center}
\begin{tabular}{c|c|c|c}
$\beta$&5.8&6.0&6.2\\\hline
$\xi_{0,4}$&3.04(3)&3.13(6)&3.29(4)
\end{tabular}
\end{center}
\end{table}

\begin{figure}
\centerline{\epsfxsize=8cm\epsfbox{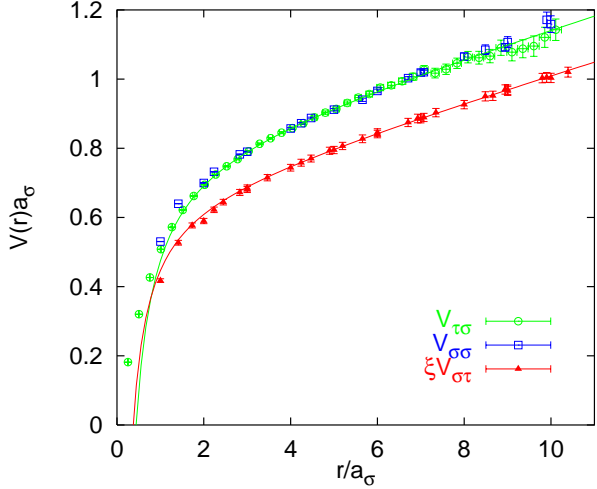}}
\caption{The three potentials $V_{\tau\sigma}$, $V_{\sigma\sigma}$
and $V_{\sigma\tau}$ in units of $a_{\sigma}$
at $\beta=6.2$, $\xi_0=3.25$.}
\label{fig:asymfix}
\end{figure}

\begin{figure}
\centerline{\epsfxsize=8cm\epsfbox{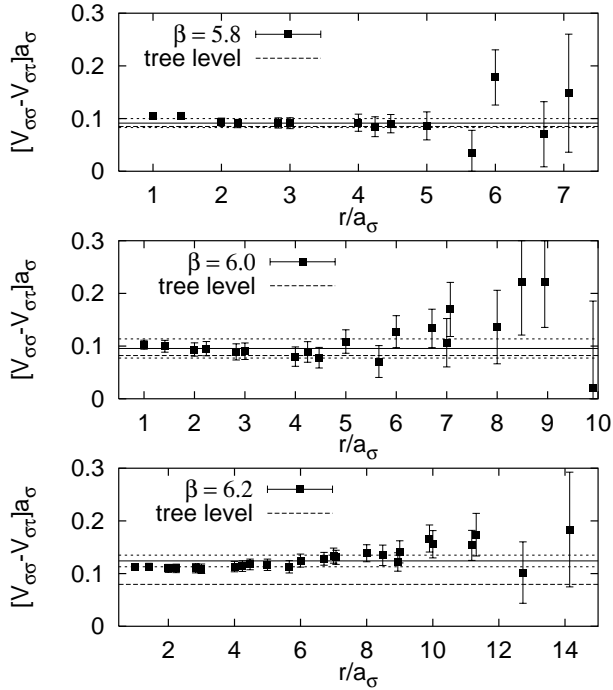}}
\caption{Differences between $\hat{V}_{\sigma\sigma}$
and  $\xi\hat{V}_{\sigma\tau}$.}
\label{fig:diff}
\end{figure}

\begin{table}[hbt]
\caption{Fits to $V_{\sigma\tau}$. The fit parameters are displayed
in units of $a_{\sigma}$.}
\label{tab:r0}
\begin{center}
\begin{tabular}{c|c||c|c|c|c|c}
$\beta$&$L_{\sigma}$&$t_{\min}/a_{\tau}$&$r_{\min}/a_{\sigma}$&
$\hat{V}_{0,\sigma\tau}\xi$&$e_{\sigma\tau}\xi$&$K_{\sigma\tau}\xi$\\\hline
5.8&8 &13&$\sqrt{5}$&0.756(50)&0.348(74)&0.138(8)\\
5.8&12&13&$\sqrt{5}$&0.732(55)&0.298(80)&0.145(9)\\
6.0&12&10--11&3&0.745(39)&0.354(80)&0.0659(41)\\
6.2&16&16--19&4&0.658(25)&0.249(67)&0.0376(17)
\end{tabular}
\end{center}
\end{table}

Assigning the phenomenological
value~\cite{Sommer:1994ce,Bali:1997am,Bali:1998pi,Bali:2000gf}
$0.5$~fm to $r_0$
we find $L_{\sigma}a_{\sigma}\approx 1.3$~fm on the small lattices and
$L_{\sigma}a_{\sigma}\approx 2$~fm on the $12^3$ lattice at $\beta=5.8$.
This means that
$\sqrt{3}L_{\sigma}a_{\sigma}/2>1.1$~fm in all our simulations;
along the
${\mathbf R}\parallel (1,1,1)$ direction we are safe from the effect
of mirror charges up to distances bigger than one fm. Beyond this
distance
only representations with non-zero triality are protected by
the centre symmetry~\cite{Bali:2000gf} from
direct finite size effects.

In Fig.~\ref{fig:asymfix} we display all three potentials in
units of $a_{\sigma}$ at $\beta=6.2$.
Note that the anisotropy has been determined by matching
$V_{\tau\sigma}$ to $V_{\sigma\sigma}$ at $r=4\,a_{\sigma}$.
In addition to the data points two curves are included
that correspond to the parameter values of Table~\ref{tab:aniso}
and Table~\ref{tab:r0} from fits according to
Eq.~(\ref{eq:fitp}) to $V_{\tau\sigma}$ for $r>2a_{\sigma}$ and to
$V_{\sigma\tau}$ for $r\geq 4a_{\sigma}$, respectively.
The matched potentials $V_{\sigma\sigma}$ and $V_{\tau\sigma}$
follow the same curve.

Up to lattice artefacts
and the self energy shift $\Delta\hat{V}_{\mbox{\scriptsize self}}$,
$\xi\hat{V}_{\sigma\tau}$ and $\hat{V}_{\sigma\sigma}$, that
both live along coarse grained lattice directions,
should also
agree with each other [Eq.~(\ref{eq:match2})].
Indeed, as is demonstrated in Fig.~\ref{fig:diff},
the differences are compatible with constants of the order suggested
by tree level perturbation theory for $\xi=\xi_0=4$, Eq.~(\ref{eq:vdiff}).
Averaging the $R\geq R_m^L$ data points
results in the
values, $0.091(9)$, $0.096(18)$ and
$0.124(11)$ for $\Delta\hat{V}_{\mbox{\scriptsize self}}$
at $\beta = 5.8,6.0$ and 6.2, respectively
(solid lines with error bands).
On the large lattice at $\beta=5.8$ we obtain the value
$0.081(9)a_{\sigma}^{-1}$, in agreement with that above,
from the smaller volume. These shifts of the self energies result
in reduced relative
errors of $\langle W_{\sigma\tau}\rangle$ [Eq.~(\ref{eq:errorr})]
(and in increased errors of $\langle W_{\sigma\sigma}\rangle$),
relative to the isotropic case.

\section{The potentials}
\label{sec:pots}
The potentials in non-fundamental representations
are extracted in the same way as discussed above from fits
to the corresponding smeared Wilson loops.
These are obtained from the fundamental ones by use of
Eqs.~(\ref{eq:reps}) -- (\ref{eq:reps15s}).
In the case of the fundamental Wilson loops, discussed
in Sec.~\ref{sec:aniso},
temporal links have been
replaced by their thermal averages in the vicinity of the surrounding
staples in order to reduce statistical
fluctuations~\cite{Bali:1997am,DeForcrand:1985dr}
(link integration~\cite{Parisi:1983hm}).
Note that our use of Eqs.~(\ref{eq:reps}) -- (\ref{eq:reps15s})
implies that we cannot
thermally average fundamental links in the construction of higher
representation Wilson loops anymore.

We determine the potentials
from correlated exponential
fits to $\langle W_{D,\sigma\tau}\rangle$ data
according to Eq.~(\ref{eq:po2}).
The fit range in $T$ is selected separately
for each distance ${\mathbf R}$ and representation $D$,
such that
$\chi^2/N_{DF}<1.5$. In addition, we demand the saturation of
``effective masses'',
$V_D(T)a_{\sigma}=
\frac{\xi}{4}\ln[\langle W_D(T)\rangle/\langle W_D(T+4)\rangle]$, into plateaus
for $T\geq T_{\min}$. In Table~\ref{tab:range} we display
the resulting
fit ranges $T\in [T_{\min},T_{\max}]$ that have been selected by means of this
procedure for the example of the point $\hat{V}_D({\mathbf R})$
with $R\approx r_0/a_{\sigma}$. In general, the interplay between
statistical errors and ground state overlaps was such that
$T_{\min}$ only slightly varied with $R$.
In the case of the fundamental potential we find values
$2\,r_0\leq t_{\min}\leq 4.5\,r_0$, depending on $R$ and the parameter values
we simulate at, while for $D={\mathbf 1}{\mathbf 5}{\mathbf s}$
we find,
$0.7\,r_0\leq t_{\min}\leq r_0$.
The corresponding estimated ground state overlaps
$c_D$ are displayed in Table~\ref{tab:overlap}.
The overlaps decrease with increasing Casimir constant,
lattice spacing or distance $r$.
At $r\approx r_0$ the overlaps range from $0.62(3)$ in the worst case to
$0.97(1)$ in the best case which quantifies the efficiency of our
smearing algorithm.

\begin{figure}
\centerline{\epsfxsize=8cm\epsfbox{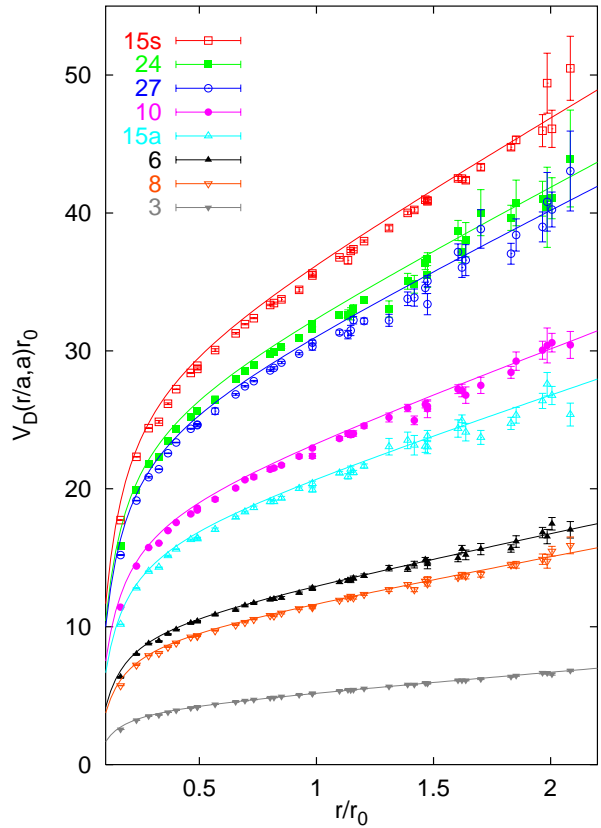}}
\caption{The potentials for all measured representations,
obtained at $\beta=6.2$. Note that
we did not subtract any self energy pieces but just rescaled
the raw lattice data in units of $r_0$.}
\label{fig:potrep}
\end{figure}

\begin{figure}
\centerline{\epsfxsize=8cm\epsfbox{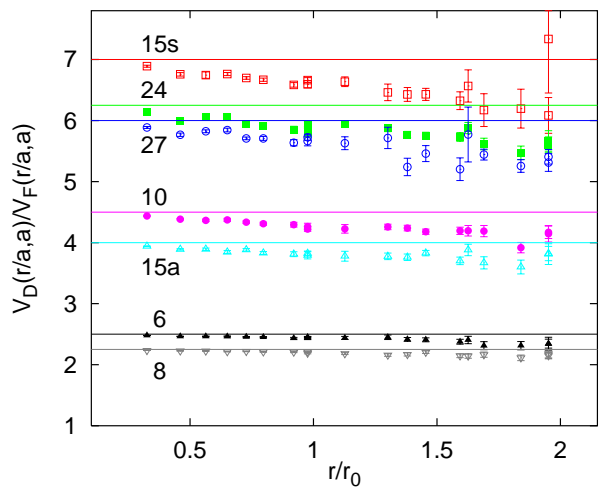}}
\caption{The potentials normalised to the fundamental
potential at $\beta=5.8$, in comparison to the expectations
from Casimir scaling (horizontal lines).}
\label{fig:ratio1}
\end{figure}

\begin{figure}
\centerline{\epsfxsize=8cm\epsfbox{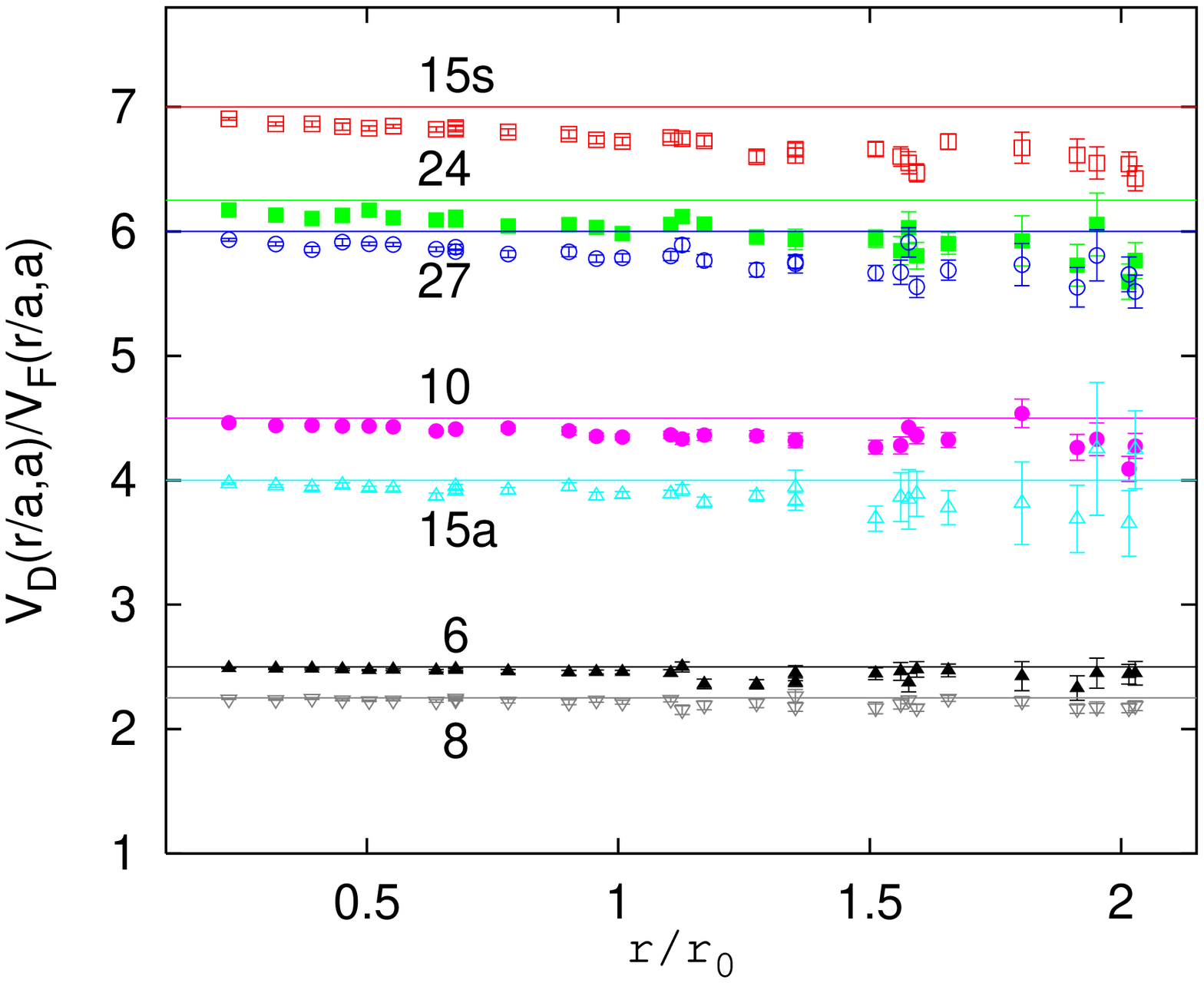}}
\caption{The same as Fig.~\ref{fig:ratio1} at $\beta=6.0$.}
\label{fig:ratio2}
\end{figure}

\begin{figure}
\centerline{\epsfxsize=8cm\epsfbox{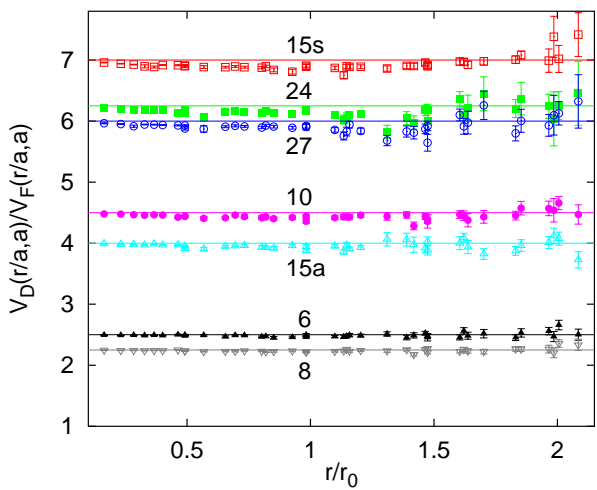}}
\caption{The same as Fig.~\ref{fig:ratio1} at $\beta=6.2$.}
\label{fig:ratio3}
\end{figure}

In Fig.~\ref{fig:potrep} we display the resulting potentials at
$\beta=6.2$. The curves correspond to the three parameter fit
Eq.~(\ref{eq:fitp}) to the fundamental potential, multiplied by
the respective ratio of Casimir factors, $d_D$ of Table~\ref{tab:reps}.
It is clear that the Casimir scaling hypothesis Eq.~(\ref{eq:pothigh})
works quite well on our $\beta=6.2$ data for the investigated distances.
In Figs.~\ref{fig:ratio1} -- \ref{fig:ratio3} 
we display the ratios $\hat{V}_D(R,a_{\sigma})/\hat{V}_F(R,a_{\sigma})$
for our three lattice spacings.
We did not attempt to subtract the self energy contributions
[cf.\ Eqs.~(\ref{eq:po1}) -- (\ref{eq:conte})] in this comparison.
At $\beta=5.8$ and $\beta=6.0$
we find the data to lie significantly below
the corresponding
Casimir ratios (horizontal lines). However, the deviations decrease
rapidly as the lattice spacing is reduced.

\begin{table}[hbt]
\caption{Fit ranges $t_{\min}/a_{\tau}$ -- $t_{\max}/a_{\tau}$
at $r\approx r_0$.}
\label{tab:range}
\begin{center}
\begin{tabular}{c|c|c|c}
$D$&$\beta=5.8$&$\beta=6.0$&$\beta=6.2$\\\hline
3  &13 -- 24&10 -- 36&16 -- 40\\
8  & 7 -- 12& 9 -- 18&10 -- 20\\
6  & 7 -- 12& 8 -- 18&10 -- 20\\
15a& 6 -- 11& 6 -- 14& 8 -- 15\\
10 & 5 -- 10& 5 -- 11& 6 -- 12\\
27 & 4 --  8& 4 --  8& 5 --  9\\
24 & 4 --  8& 4 --  8& 5 --  9\\
15s& 3 --  7& 3 --  7& 4 --  7
\end{tabular}
\end{center}
\end{table}

\begin{table}[hbt]
\caption{Ground state overlaps $c_D$ at $r\approx r_0$.}
\label{tab:overlap}
\begin{center}
\begin{tabular}{c|c|c|c}
$D$&$\beta=5.8$&$\beta=6.0$&$\beta=6.2$\\\hline
3  &0.89 (1)&0.96 (1)&0.97 (1)\\
8  &0.81 (1)&0.90 (3)&0.91 (2)\\
6  &0.80 (2)&0.90 (3)&0.90 (3)\\
15a&0.68 (4)&0.84 (3)&0.88 (5)\\
10 &0.62 (3)&0.82 (2)&0.89 (3)\\
27 &0.67 (3)&0.81 (2)&0.88 (3)\\
24 &0.65 (3)&0.77 (2)&0.90 (4)\\
15s&0.66 (3)&0.79 (2)&0.86 (3)
\end{tabular}
\end{center}
\end{table}

Prior to a continuum limit extrapolation of the ratios
we shall
investigate finite size effects by comparing results obtained
on the 1.3~fm lattice with results from the 2~fm lattice at $\beta=5.8$.
For the
fundamental potential we already know from previous studies that
for spatial extents, $L_{\sigma}a_{\sigma}>2\,r_0$,
such effects are well below the 2~\% level~\cite{Bali:1992ab}.
The situation is less clear for potentials in higher representations.
In principle, the flux tube between the sources could widen when
the energy per unit length
is increased and, therefore, higher representation
potentials might be more susceptible to finite size effects.

\begin{figure}
\centerline{\epsfxsize=8cm\epsfbox{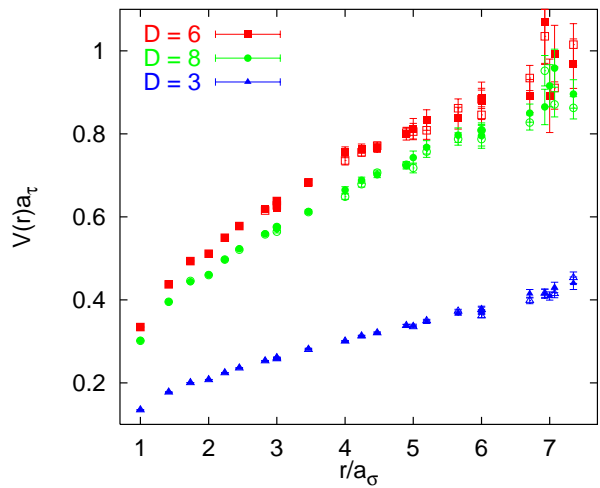}}
\caption{Comparison between the potentials in the lowest dimensional
three representations between the $L_{\sigma} = 8$ lattices
(open symbols) and $L_{\sigma} = 12$ lattices (full symbols)
at $\beta = 5.8$ in lattice units.}
\label{fig:fse}
\end{figure}

\begin{figure}
\centerline{\epsfxsize=8cm\epsfbox{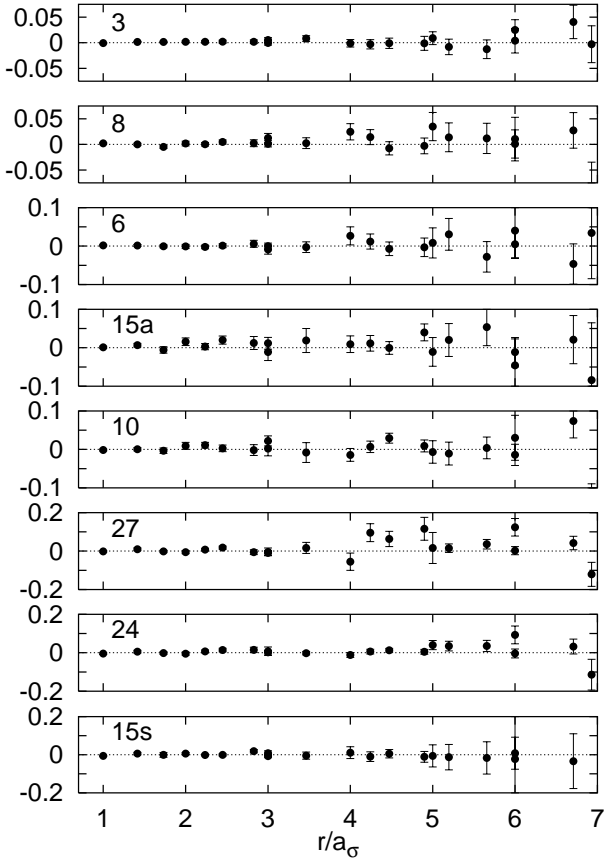}}
\caption{Relative deviations $V^{L_{\sigma}=12}_D({\mathbf r})/
V^{L_{\sigma}=8}_D({\mathbf r})-1$
of the potentials obtained on different volumes as a function
of $r/a_{\sigma}$.}
\label{fig:difffse}
\end{figure}

In Fig.~\ref{fig:fse} we compare the fundamental, octet and sextet
potentials obtained on the $12^3$
lattices at $\beta=5.8$ (full symbols) with those obtained on the
$8^3$ lattices (open symbols).
Up to distances well beyond $2\,r_0\approx 6\,a_{\sigma}$
no statistically significant deviations are seen.
In Fig.~\ref{fig:difffse} we show the relative deviations,
$V^{L_{\sigma}=12}_D({\mathbf r})/V^{L_{\sigma}=8}_D({\mathbf r})-1$,
between the potentials
determined on the larger lattices and those measured
on the smaller lattices
for all the representations that we have investigated.
Again, no systematic or statistically significant differences are
detected. Up to $r=4\,a_{\sigma}\approx 1.4\,r_0$ this holds true
on the 1~\% level for the fundamental potential and
on the 3--5~\% level for higher representation potentials.
Beyond this distance the statistical errors start to explode.
The same comparison has been performed for ratios of potentials.
In this case the relative errors are slightly reduced due to
correlation effects. However, no statistically significant
tendencies were observed either.
The relative statistical errors on the two lattices are of about
the same size and comparable to those of the $\beta=6.0$
and $\beta=6.2$ simulations. Thus, we do not expect
finite size effects to exceed the statistical errors
on any of the simulated lattice volumes.

We now attempt a continuum limit extrapolation of our data.
We remark that in the limit $a_{\sigma}\rightarrow 0$
[Eqs.~(\ref{eq:po1}) -- (\ref{eq:conte})] the Casimir scaling of
the diverging
self energies $\hat{V}_{D,\mbox{\scriptsize self}}(a_{\sigma})$
automatically implies Casimir scaling of
$\hat{V}_D({\mathbf R},a_{\sigma})$.
However, with $V_{D,\mbox{\scriptsize self}}$ being a purely ultra violet
quantity, this sort of Casimir scaling
has little to do with non-perturbative aspects of the
theory.
As can be seen from Fig.~\ref{fig:potrep}, where the potentials
vary by more than a factor two with the distance, even at our
finest lattice resolution
$V_{D,\mbox{\scriptsize self}}$ have not yet become the dominant
contributions to the lattice potentials.
In order to avoid the
trivial Casimir scaling described above
we will only study
ratios of physical interaction energies,
\begin{equation}
\label{eq:extra}
R_D(r)=\frac{V_D(r)}{V_F(r)}=\frac{\hat{V}_D({\mathbf R},a_{\sigma})
-\hat{V}_{D,\mbox{\scriptsize self}}(a_{\sigma})}
{\hat{V}_F({\mathbf R},a_{\sigma})
-\hat{V}_{F,\mbox{\scriptsize self}}(a_{\sigma})}
\left[1+{\mathcal O}(a_{\sigma}^2)\right],
\end{equation}
where $r=Ra_{\sigma}$.

We estimate the self energies in leading order
perturbation theory, Eq.~(\ref{eq:selfen}).
Lattice perturbation theory is notorious
for its bad convergence behaviour~\cite{Bali:1993ru,Lepage:1993xa}.
However, in our anisotropic case the effective expansion parameter
$g^2\xi_0^{-1}$ is much smaller than
in standard applications of lattice perturbation theory. 
In Fig.~\ref{fig:diff} we have indeed
seen that leading order perturbation theory predictions
on the difference of two self energies agree
within 30~\% with numerical data. We estimate the self energies
in two different ways:
(a) we use the bare lattice coupling and the
renormalised anisotropy $\xi$, (b)
we mean field (``tadpole'')
improve~\cite{Parisi:1980pe,Lepage:1993xa} both, coupling
and anisotropy [Eq.~(\ref{eq:tadpole})],
$\xi_{0,ir}^{-1}g^2_{ir}=\xi_0^{-1}g^2\langle U_{\sigma\tau}\rangle^{-1}$.
The results from the two methods,
shown in Table~\ref{tab:vself}, differ
by up to 60~\% from each other.
We will use the estimate (a) in our analysis but
take the difference between (a) and (b) into account as a systematic error.
While data at large distances are marginally affected by this uncertainty,
the error bars at small distances are significantly increased.

\begin{figure}
\centerline{\epsfxsize=8cm\epsfbox{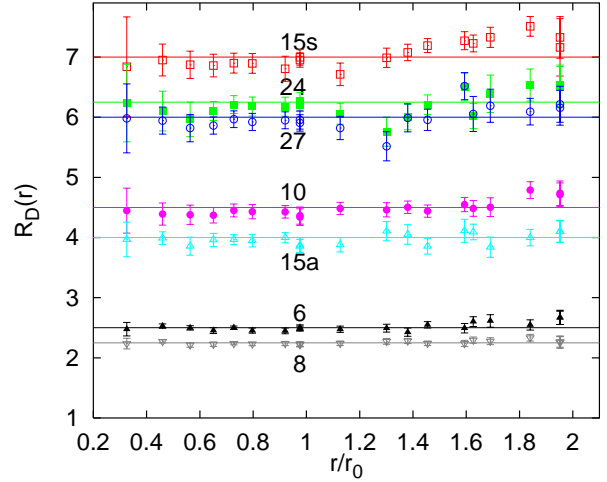}}
\caption{Continuum limit extrapolated ratios, Eq.~(\protect\ref{eq:extra}),
compared to
the Casimir scaling hypothesis (horizontal lines).}
\label{fig:ratioe}
\end{figure}

After subtracting the (scaling violating) self energy contributions
we determine the continuum extrapolated ratios $R_D(r)$
by means of quadratic fits, Eq.~(\ref{eq:extra}). We perform
these extrapolations for all the distances $r$ that
have been realised on our coarsest lattice ($\beta=5.8$) in units
of $r_0$.
On the finer lattices, we linearly interpolate between
the two lattice points that
are closest to each given distance $r$, prior to
the quadratic continuum limit fit. We find the data to be compatible with
the quadratic ansatz and the
resulting ratios are shown in Fig.~\ref{fig:ratioe}. The
numerical values are displayed in Tables~\ref{tab:rat1} -- \ref{tab:rat2}.
No statistically significant violations of Casimir scaling are found.
Our accuracy is somewhat limited at short distances, due to the
perturbative estimation of the self energies.
The slope of the extrapolation in $a_{\sigma}^2$ increases
with the distance $r$ as well as with the Casimir charge $C_D$;
large masses $\hat{V}_D(R,a_{\sigma})>a_{\tau}^{-1}$ are more
affected by lattice artefacts than small masses. This observation
also explains why the deviations from Casimir scaling at $\beta=5.8$ and
$\beta=6.0$
increase with the distance (Figs.~\ref{fig:ratio1} -- \ref{fig:ratio2}).

\begin{table}
\caption{Perturbative estimates
of the self energy $\hat{V}_{F,\mbox{\scriptsize self}}$
(in units of $a_{\tau}$).}
\label{tab:vself}
\begin{center}
\begin{tabular}{c||c|c}
$\beta$&bare $g^2$, renormalised $\xi$
&mean field $g_{ir}^2\xi_{0,ir}^{-1}$\\\hline
5.8&0.0855(05)&0.13931\\
6.0&0.0825(15)&0.12835\\
6.2&0.0824(11)&0.12093
\end{tabular}
\end{center}
\end{table}

\begin{table}[hbt]
\caption{The continuum ratios, \protect{$R_D(r)=V_D(r)/V_F(r)$.}
``exp.'' denotes the Casimir scaling expectation.}
\label{tab:rat1}
\begin{center}
\begin{tabular}{c|c|c|c}
$r/r_0$&$D=8$&$D=6$&$D=15a$\\\hline
0.33&2.24(09)&2.48(11)&3.97(29)\\
0.46&2.27(03)&2.53(04)&3.99(10)\\
0.56&2.21(04)&2.50(04)&3.86(15)\\
0.65&2.23(04)&2.45(05)&3.96(11)\\
0.73&2.24(02)&2.50(03)&3.97(08)\\
0.80&2.23(03)&2.45(04)&3.95(11)\\
0.92&2.24(03)&2.45(05)&4.00(08)\\
0.98&2.23(04)&2.50(05)&3.87(12)\\
0.98&2.22(04)&2.49(04)&3.85(11)\\
1.13&2.24(04)&2.48(05)&3.88(12)\\
1.30&2.27(05)&2.49(07)&4.11(16)\\
1.38&2.28(04)&2.43(07)&4.05(16)\\
1.45&2.24(04)&2.55(05)&3.86(13)\\
1.59&2.24(05)&2.49(08)&4.11(19)\\
1.63&2.29(05)&2.60(08)&4.09(13)\\
1.69&2.28(06)&2.61(10)&3.84(17)\\
1.84&2.33(06)&2.55(09)&4.00(14)\\
1.95&2.26(10)&2.67(12)&4.10(19)\\
1.95&2.27(09)&2.66(11)&4.10(18)\\
2.18&2.30(11)&2.11(16)&3.67(26)\\
2.25&1.97(16)&2.03(20)&3.41(43)\\
2.30&2.26(14)&2.45(17)&3.72(31)\\
2.39&2.29(14)&2.46(16)&\\\hline
exp.&2.25&2.5&4
\end{tabular}
\end{center}
\end{table}

\begin{table}[hbt]
\caption{The same as Table~\ref{tab:rat1} for higher representations.}
\label{tab:rat2}
\begin{center}
\begin{tabular}{c|c|c|c|c}
$r/r_0$&$D=10$&$D=27$&$D=24$&$D=15s$\\\hline
0.33&4.45(37)&6.23(65)&5.98(57)&6.84(83)\\
0.46&4.39(18)&6.10(33)&5.94(22)&6.95(27)\\
0.56&4.38(16)&5.96(29)&5.82(23)&6.87(23)\\
0.65&4.37(13)&6.10(16)&5.86(14)&6.86(19)\\
0.73&4.45(11)&6.21(15)&5.97(14)&6.90(17)\\
0.80&4.43(12)&6.18(16)&5.92(14)&6.89(17)\\
0.92&4.43(10)&6.17(16)&5.95(14)&6.81(21)\\
0.98&4.34(14)&6.28(14)&5.95(15)&7.00(13)\\
0.98&4.37(14)&6.21(12)&5.91(14)&6.95(12)\\
1.13&4.48(10)&6.06(17)&5.82(19)&6.71(19)\\
1.30&4.46(11)&5.76(24)&5.52(24)&6.99(16)\\
1.38&4.50(10)&6.00(21)&5.99(23)&7.08(14)\\
1.45&4.44(10)&6.21(16)&5.96(18)&7.19(12)\\
1.59&4.55(12)&6.50(23)&6.52(23)&7.27(15)\\
1.63&4.48(13)&6.02(21)&6.05(29)&7.23(14)\\
1.69&4.50(16)&6.39(31)&6.19(28)&7.33(17)\\
1.84&4.79(14)&6.53(27)&6.09(22)&7.51(16)\\
1.95&4.74(20)&6.51(33)&6.21(30)&7.16(48)\\
1.95&4.71(19)&6.54(32)&6.16(29)&7.33(35)\\
2.18&4.40(36)&5.93(53)&5.89(43)&7.19(57)\\
2.25&3.12(50)&5.67(71)&5.99(58)&\\
2.30&3.82(44)&6.37(76)&5.97(53)&\\
2.39&3.62(53)&        &        &\\\hline
exp.&4.5&6&6.25&7
\end{tabular}
\end{center}
\end{table}

\section{Discussion}
We have confirmed that
violations of the Casimir scaling hypothesis,
\begin{equation}
\frac{V_{D_1}(r)}{V_{D_2}(r)}
=\frac{C_{D_1}}{C_{D_2}},
\end{equation}
are below the 5~\% level for distances $r<2\,r_0\approx 1$~fm
in the continuum limit of four dimensional
$SU(3)$ gauge theory for all representations with
$C_D\leq 7$.
This finding rules out many models of non-perturbative QCD
and imposes serious
restrictions onto others. For instance
in a bag model calculation scaling of string tensions
with the square root of the respective Casimir ratio
has been obtained~\cite{Johnson:1976sg} and instanton
liquid calculations result in
ratios between potentials in different
representations that are smaller than the Casimir ratios
too~\cite{Diakonov:1989un}.

Another possibility would have been
scaling proportional to the number
of fundamental flux tubes embedded into the higher representation
vortex [$p+q$ in $SU(3)$]~\cite{Michael:1998sm,Bali:2000hx}, which
happens to coincide with Casimir scaling in the large $N$ limit of $SU(N)$.
This picture is supported by the finding that the $SU(N)$ vacuum
seems to act like a type I superconductor~\cite{Bali:1997cp,Bali:1998de},
i.e.\
flux tubes repel each other.
However, this scenario is also excluded by the present study.
Furthermore, serious restrictions onto most of the remaining models
are imposed (see for instance Ref.~\cite{Shevchenko:2000du}).
It is particularly disappointing that neither
centre vortex models~\cite{Faber:1998rp,Cornwall:1998ds,Deldar:1999yy}
nor the dual superconductor
scenario~\cite{Bali:1996dm,Ambjorn:1998qp,Bali:1998de}
or string models~\cite{Luscher:1981iy}
seem to offer any explanation why the numerical data
so closely resemble the Casimir ratios.
Certainly, it is worthwhile to dedicate
more theoretical effort to
this fundamental phenomenon.

We have not discussed ``string breaking'' so far.
While the fundamental potential in pure gauge theories linearly
rises {\em ad infinitum}, the adjoint potential
will be screened by gluons and,
at sufficiently large distances, decay into two disjoint
gluelumps~\cite{Jorysz:1988qj,Michael:1998sm,Bali:2000gf}.
This string breaking has indeed been
confirmed in numerical
studies~\cite{Markum:1988na,Muller:1991xj,Stephenson:1999kh,Philipsen:1999wf,deForcrand:1999kr}. Therefore,
strictly speaking, the adjoint string tension is {\em zero}. 
In fact, all charges in higher than the fundamental representation
will be at least partially screened by the background gluons. For instance,
${\mathbf 6}\otimes {\mathbf 8}={\mathbf 2}{\mathbf 4}\oplus
{\mathbf 1}{\mathbf 5}{\mathbf a}^*\oplus {\mathbf 6}\oplus {\mathbf 3}^*$:
in interacting with the glue, the sextet potential obtains a fundamental
component. 
A simple rule, related to the centre of
the group, is reflected in Eqs.~(\ref{eq:reps}) -- (\ref{eq:reps15s}):
wherever $z^{p-q}=1$ (zero triality),
the source will be reduced into a singlet component
at large distances while, wherever
$z^{p-q}=z$ (or $z^*$), it will
be screened, up to a residual (anti-)\-triplet
component, i.e.\
one can easily read off the asymptotic string tension (either {\em zero} or the
fundamental string tension) from
the third column of Table~\ref{tab:reps}, rather than having to multiply
and reduce representations. As a result, the self-adjoint representations,
${\mathbf 8}$ and ${\mathbf 2}{\mathbf 7}$, as well as the representation,
${\mathbf 1}{\mathbf 0}$, will be completely
screened while in all other representations with $p+q\leq 4$ a residual
fundamental component survives. The same argument, applied to
$SU(2)$, results in the prediction that all odd-dimensional (bosonic)
representations
are completely screened while all even-dimensional (fermionic)
representations will tend towards the fundamental string tension at large
distances.

One expects this sort of string breaking and flattening of the potential
to occur at distances larger than about 2.4~$r_0$~\cite{Michael:1998sm}.
Obviously, once the string is broken Casimir scaling is violated.
It is certainly interesting to investigate what happens around the
string breaking distance. However, this requires lattice volumes exceeding
those used in the present study as well as additional operators
that are designed for an optimal overlap with the
respective broken string states~\cite{Schilling:2000mv}.

\acknowledgements
This work was supported by DFG grants
Ba 1564/3-1, 1564/3-2 and 1564/3-3 as well as
EU grant HPMF-CT-1999-00353. The simulations were
performed on the Cray J90 system of the ZAM 
at Forschungszentrum J\"ulich as
well as on workstations of the John von Neumann Institut f\"ur Computing.
We thank the support teams of these institutions for their help.

\end{document}